\documentclass[twocolumn,floatfix,amssymb,aps,prb,nobibnotes,nofootinbib]{revtex4-2}
\usepackage{docs}
\usepackage{xcolor}
\usepackage{graphicx}
\usepackage{float}
\usepackage{amsmath}
\usepackage{pifont}

\expandafter\ifx\csname package@font\endcsname\relax\else
 \expandafter\expandafter
 \expandafter\usepackage
 \expandafter\expandafter
 \expandafter{\csname package@font\endcsname}%
\fi
\DeclareRobustCommand\substyle{\name@idx{document substyle}}%
\DeclareRobustCommand\classoption{\name@idx{document class option}}%
\DeclareRobustCommand\classname{\name@idx{document class}}%
\def\name@idx#1#2{%
 {\ttfamily#2}%
 \index{#2\space#1=\string\ttt{#2}\space#1}\index{#1>#2=\string\ttt{#2}}%
}%

\begin{document}
\title{Measurement-Driven Transitions between Area Law Phases}
\author{Hui Yu$^{1}$}
\email{yuhui252@iphy.ac.cn}
\author{Jiangping Hu$^{1,2,3}$}
\affiliation{$^{1}$Beijing National Laboratory for Condensed Matter Physics and Institute of Physics, Chinese Academy of Sciences, Beijing 100190, China}
\affiliation{$^{2}$Kavli Institute of Theoretical Sciences, University of Chinese Academy of Sciences, Beijing 100190, China}
\affiliation{$^{3}$New Cornerstone Science Laboratory, Beijing 100190, China}

\date{\today}%


\begin{abstract}
In recent years, quantum circuits consisting of unitary gates and projective measurements have become valuable tools for stimulating or preparing quantum many-body states with non-trivial properties. Here, we introduce and examine a measurement-only circuit (the projective quantum Ising model with three-spin interactions) that involves three non-commuting projective measurements. This model features three distinct phases, separated by two critical lines. We utilize two entanglement measures (topological entanglement entropy and mutual information) to identify the phase boundaries and derive various critical exponents through scaling analysis. We establish a relationship between our model and a two-dimensional statistical model (bond percolation) within certain limits. We hope that our results will shed light on further studies using other measurement-only models.  
\end{abstract}
\maketitle

\section{Introduction}
Preparing quantum states with specific properties, such as long-range or topological order, is crucial in modern physics. These states enable researchers to explore complex quantum materials and study novel phases of matter, thereby providing insights into strongly correlated systems. Recently-developed digital quantum simulators~\cite{ref1} have emerged as promising candidates for investigating the interesting phases of quantum many-body systems. These quantum processors can be viewed as quantum circuits~\cite{ref2} in which a series of quantum operations are applied to qubits in a discrete-time manner. A typical quantum circuit comprises unitary gates, projective measurements, and feedback mechanisms. 
Recently, a one-dimensional simplified circuit model has been proposed that consists of random unitary gates arranged in a brick-wall fashion, interspersed with projective measurements. Researchers have found that this model undergoes an entanglement phase transition when the measurement probability varies. This phase transition, driven by  competition between random unitary gates and projective measurements, is referred to as measurement-induced phase transition (MIPT)~\cite{ref3,ref4,ref5,ref6} or purification phase transition~\cite{ref7,ref8}. When the circuit is subjected to only a few measurements, the unitary dynamics drive the system into a highly entangled state, which is characterized by a volume law scaling of the subsystem entanglement entropy ($S_{A} \propto L_{A}$). In contrast, when projective measurements dominate the process, the subsystem entanglement entropy adheres to the area law ($S_{A} \propto Constant$). 

\begin{figure}[htb]
\includegraphics[scale=0.65]{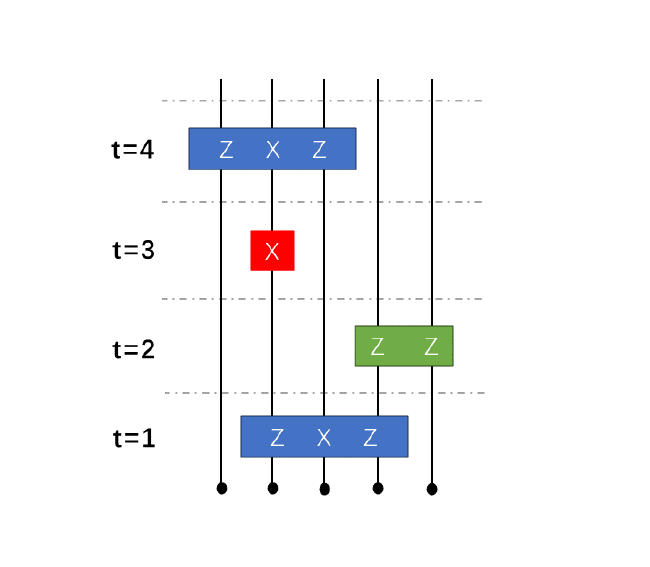}
\caption{Schematic diagram of a quantum circuit configuration: The drawing illustrates the evolution of a system with 5 qubits (black dots) over the first 4 updating steps. The red, green, and blue boxes represent different types of projective measurements. Red - Measurement $X$, Green - Measurement $ZZ$, Blue - Measurement $ZXZ$.}
\label{fig:circuit diagram}
\end{figure}

Following this, numerous studies have been conducted in this area.  From a theoretical perspective, the phase diagrams and critical properties of quantum circuits have been extensively explored under various generalizations. These include circuits with Abelian symmetries~\cite{ref9,ref10,ref11}, non-Abelian symmetries~\cite{ref12}, as well as the effects of boundary noise~\cite{ref13} and bulk noise~\cite{ref14,ref15,ref16}, all of which have been investigated numerically. Furthermore, MIPT in monitored fermion chains has been a focus of several studies, revealing rich entanglement dynamics and critical phenomena~\cite{ref17,ref18,ref19,ref20,ref21}. The connection between MIPT and conformal field theory has also been established, providing insights into the universal critical behavior and scaling properties of these transitions~\cite{ref22,ref23,ref24}.  Additionally, the relationship between certain quantum circuit models and statistical models~\cite{ref25,ref26,ref27,ref28} has been clarified. On the experimental front, MIPT signatures have been successfully observed in small-scale trapped-ion (or superconducting) quantum computers~\cite{ref29,ref30}, despite the challenges related to post-selection~\cite{ref31}.  

\begin{figure}[htbp]
\begin{center}
\includegraphics[scale=0.285]{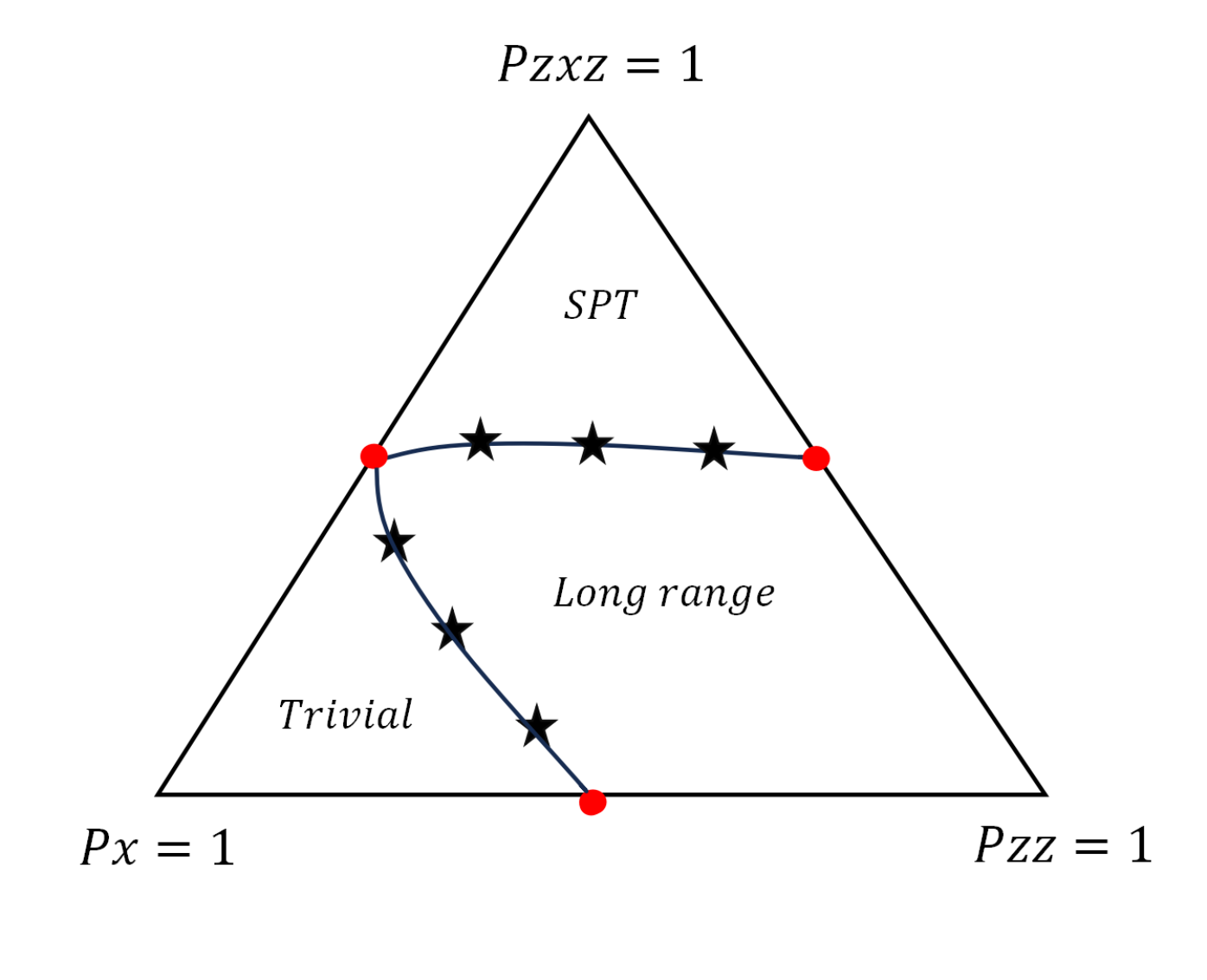}
\caption{Phase diagram of our quantum circuit, where $P_{X}$, $P_{ZZ}$, and $P_{ZXZ}$ serve as tuning parameters constrained by $P_{X}+P_{ZZ}+P_{ZXZ}=1$. The phases correspond to distinct entanglement structures in the steady state. The SPT phase is characterized by topological order. The Long-range phase by long range entanglement, and the Trivial phase by the absence of entanglement. Red circles and black stars denote critical points that are obtained through numerical simulations, while the remaining phase boundaries are extrapolated.}
\label{fig:Phase Diagram}
\end{center}
\end{figure}

In this paper, we focus on a novel type of quantum circuit driven solely by random projective measurements. It is well established that states with various orders can emerge in quantum systems with multiple dissipative channels~\cite{ref32,ref33}. Previous studies ~\cite{ref34,ref35} have examined the properties of generic measurement-only models and their mappings to Majorana fermions. The circuit we consider employs a specific set of measurement gates: $X$, $ZZ$, and $ZXZ$. We refer to this model as the projective quantum ising model with three-spin interactions, as these measurement gates correspond to terms in the following Hamiltonian~\cite{ref39,ref40,ref41}:
\begin{equation}
H=-\sum_i(h \sigma_i^x+\lambda_2\sigma_i^{x}\sigma_{i-1}^z\sigma_{i+1}^z+\lambda_1\sigma_i^z\sigma_{i-1}^z)
\label{Eq:Hamilitonian}
\end{equation}
Our study of this circuit model is motivated by two key questions: (1) whether its phase diagram shares structural similarities with that of the Hamiltonian system, and (2) whether its critical behavior aligns with known statistical models, such as bond percolation. The Hamiltonian Eq.~\ref{Eq:Hamilitonian} exhibits criticality in two parameter regimes: when $h=\lambda_{1}$, with $\lambda_{2}=0$, and when $h=\lambda_{2}$, with $\lambda_{1}=0$.  Recent work has revealed analogous entanglement phase transition in measurement circuits: when $P_{X}=P_{ZZ}$ with $P_{ZXZ}=0$ ~\cite{ref36,ref37}, and when $P_{X}=P_{ZXZ}$ with $P_{ZZ}=0$ ~\cite{ref38}. Our analysis confirms that the model indeed features a non-trivial phase diagram, as shown in Fig.~\ref{fig:Phase Diagram}. This diagram contains three critical points situated on its boundaries and two critical lines that separate the three distinct phases. In the trivial phase, the state of the system resembles that of the product state with no entanglement. In the long-range phase, the state possesses a long-range order, characterized by a finite amount of entanglement between distant qubits. In the symmetry protected topological (SPT) phase, the state acquires a non-trivial topological order. Various critical exponents have been obtained in this study. 

The remainder of this paper is organized as follows. In Section II, we introduce the circuit model and define the two entanglement measures used to identify the phase boundaries. Section III provides a brief discussion of the methods employed in this study. In Section IV, the numerical results are presented and analyzed. Section V establishes the connection between proposed circuit and  bond percolation model. Finally, Sec VI concludes with a short summary.

\section{The Measurement-only Model}
We begin by providing a detailed description of the circuit model. Our model consists of a one-dimensional chain with $L$ sites, where each site $i$ hosts a qubit (spin-1/2 degrees of freedom). The system is treated under open boundary conditions. Initially, all qubits are set to the spin-up state, represented by the wavefunction  $|0\rangle^{\otimes L}$. For the notation, we denote $X$ and $Z$ as the usual Pauli matrices $\sigma_{x}$ and $\sigma_{z}$.

In each update step, we have three options. We either: \textbf{(1)} apply a single qubit measurement gate $X$ on qubit $i$ with probability $P_{X}$, where $i$ is drawn uniformly from $1,2,...,L$, or \textbf{(2)} apply a two-qubit measurement gate
$ZZ$ on a pair of qubits $(i,i+1)$ with probability $P_{ZZ}$, where $i$ is drawn uniformly from $1,...,L-1$, or \textbf{(3)} apply a three-qubit measurement gate $ZXZ$ between qubits $i-1$, $i$,and $i+1$ with probability $P_{ZXZ}$, where $i$ is drawn uniformly from $2,...,L-1$. A schematic drawing of the circuit diagram is provided in Fig.~\ref{fig:circuit diagram}. We define a single time step $t$ as $L$ updating steps and require $P_{x}+P_{ZZ}+P_{ZXZ}=1$. We then evolve the system iterative until a steady state is reached, typically requiring $O(L)$ time steps.

Unlike in the original works, our quantum circuit does not involve random unitary gates. However, this simplification does not trivialize the quantum dynamics of our circuit. Notably, any two of measurement operators do not commute if they involve $X$ and $Z$ on the same site. For example, $X_{i}$ does not commute with $Z_{i}Z_{i+1}$ or $Z_{i}X_{i+1}Z_{i+2}$. This non-commutativity leads to a nontrivial phase diagram with several distinct phases, as illustrated in Fig.~\ref{fig:Phase Diagram}.

\begin{figure}
\begin{center}
\includegraphics[scale=0.65]{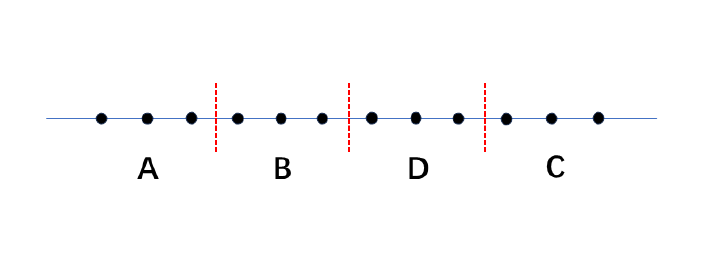}
\caption{The partition of one-dimensional chain ($L=12$) into regions $A$, $B$, $D$, and $C$, each containing an equal number of qubits, is used to define the topological entanglement entropy $S_{topo}$.} 
\label{fig:Partition}
\end{center}
\end{figure}

To detect the presence of various phases across different regions of the parameter space, we utilize two entanglement measures: topological entanglement entropy $S_{topo}$ and mutual information $I_{AB}$. In our case, the topological entanglement entropy serves as an order parameter for the SPT phase. For a one-dimensional open chain, we divide the system as shown in Fig.~\ref{fig:Partition}. The topological entanglement entropy $S_{topo}$~\cite{ref42,ref43} is defined as follows:
\begin{equation}
S_{topo} \equiv S_{AB} + S_{BC} - S_{B} - S_{ABC}
\end{equation}
Here, $S_{AB}$ represents the standard Von Neumann entanglement entropy of subsystem $A\cup B$, defined as:
\begin{equation}
S_{AB} \equiv - \text{Tr} \left[ \rho_{AB} \log_2 \rho_{AB} \right]
\end{equation}
where $\rho_{AB}$ is the reduced density matrix of the subsystem. $S_{topo}$ will yield an integer value if the system acquires a certain topological order; otherwise, it vanishes in a topologically trivial phase. 

The mutual information $I_{AB}$ between subsystems $A$ and $B$ is defined as
\begin{equation}
I(A,B) \equiv S_{A} + S_{B} - S_{AB}
\end{equation}
with all terms defined similarly. The value of $I(A,B)$ characterizes the entanglement between subsystems $A$ and $B$. $I(A,B)=0$ indicates that  subsystem $A$ is completely disentangled from $B$. On the other hand, $I(A,B)=1$ indicates that  qubits in subsystem $A$ form a bell pair with those in subsystem $B$ (i.e.,$|0_{A}0_{B}\rangle + |1_{A}1_{B}\rangle$).

The quantum dynamics of our model are described by a stochastic process of applying three types of measurement gates on a line of qubits. To fully characterize the entanglement structure of the wavefunction, physical quantities must to be averaged over numerous circuit configurations (varying locations of the measurement gates and outcomes). This averaging is performed as follows:
\begin{equation}
O \equiv \frac{1}{N}\sum_{i=1}^{N} O(\rho_{i})
\end{equation}
where $O\in[S_{topo},I_{AB}]$, $N$ are the total number of realizations, and $\rho_{i}$ is the density matrix corresponding to $i^{th}$ sample.

\begin{figure}[htbp]
\begin{center}
\includegraphics[scale=0.42]{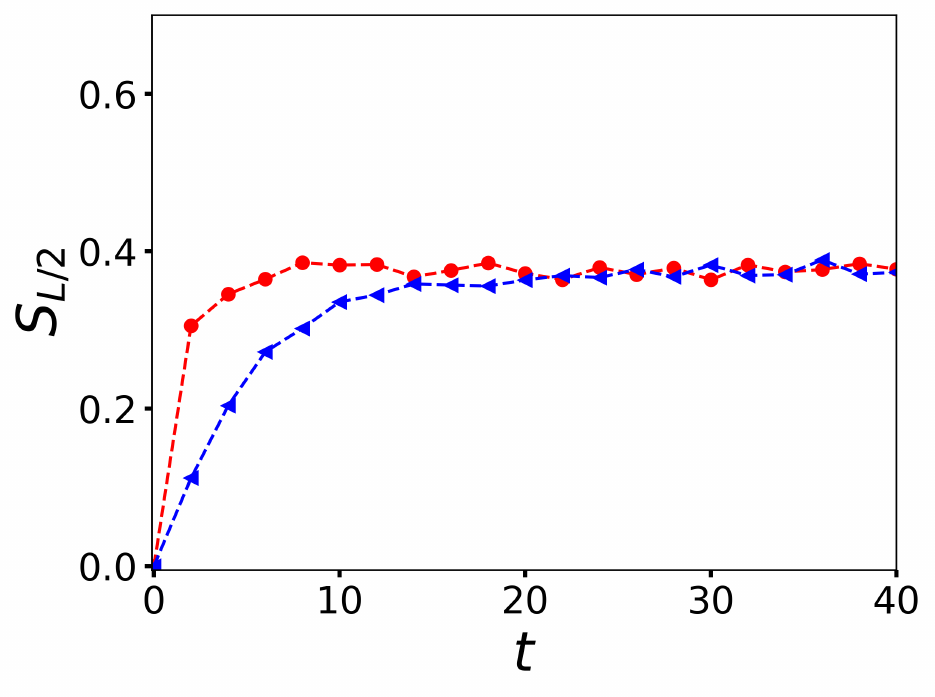}
\caption{The half-chain Von Neumann entanglement entropy $S_{L/2}$ as a function of time step $t$ for different initial states at $L=128$ with $P_{x}=P_{ZZ}=P_{ZXZ}=1/3$. The red line corresponds to $|+\rangle^{\otimes L}$, while the blue line corresponds to $|0\rangle^{\otimes L}$.}
\label{fig:EE initial state}
\end{center}
\end{figure}

\section{Methods}
In this section, we present all the methods used in our study. 
\subsection{binary representation of stabilizer cirucit}
Storing the wavefunction of a quantum circuit with hundreds of qubits as a vector is challenging due to the exponential growth of the state space, which requires $2^{n}$ complex amplitudes for $n$ qubits. This rapid increase results in significant memory constraints and computational complexity, making direct simulation impractical. However, our circuit belongs to the class of Clifford circuits~\cite{ref44}, which can be simulated efficiently in polynomial time on a classical computer, even with hundreds of qubits, due to stabilizer formalism.

In this work, we use the binary representation of stabilizer formalism to stimulate our quantum circuit. This representation was first shown and discussed in a prior paper~\cite{ref45}. Any Pauli string operator $S$ (up to a global phase factor) in a system of size $L$ can be mapped to a binary vector $w_{S}=(m_{1},m_{2},...,m_{L},n_{1},n_{2},...,n_{L})$ that has $2L$ components. The mapping is represented as:
\begin{equation}
S = \prod_{i=1}^{L}X_{i}^{m_{i}}\prod_{j=1}^{L}Z_{j}^{n_{j}}
\end{equation}
where the values $m_{i}$ and $n_{j}$ belong to $Z_{2}$. From this point onward, all manipulations are carried out over $Z_{2}$. To check the commutation relation between Pauli string operators $S_{1}$ and $S_{2}$, one can add their binary vectors $w_{S_{1}}$ and $w_{S_{2}}$. If the resulting vector contains even number of ones, $S_{1}$ and $S_{2}$ are commute; otherwise they anticommute.

In the stabilizer formalism, a stabilizer state $|\psi\rangle$ over $L$ qubits is specified by a set $T=[S_{1},...S_{L}]$ consisting of $L$ independent, mutually commuting Pauli string operators $S_{i}$, each satisfying the following property:
\begin{equation}
S_{i}|\psi\rangle = |\psi\rangle
\end{equation}
With the help of this mapping, one can construct a $L\times 2L$ stabilizer matrix $M_{T}$ by taking the binary representation of the elements of $T$ as its rows. For example, a stabilizer matrix for the state $|+\rangle^{\otimes L}$ is given by:
\begin{equation}
M_{T} = \left(1_{L \times L} \big| 0_{L \times L} \right)
\end{equation}
where $1_{L \times L}$ is an $L\times L$ identity matrix.

In our stimulation, the state $|\psi\rangle$ updates at each step when a measurement is performed and so does $M_{T}$. The updating rules for $M_{T}$ are as follows: Suppose we have a stabilizer set $T=[S_{1},...S_{L}]$ at a specific time step, and a measurement with the corresponding Pauli string operator $S^{*}$ is applied to the circuit at the next time step. We then encounter two cases. \textbf{Case 1:} If $S^{*}$ commutes with each element in the stabilizer set $T$, then no updates are needed for $M_{T}$ since this measurement has no effect on the state. \textbf{Case 2:} If $S^{*}$ does not commute (or anticommutes) with some elements (for example $S_{1},S_{2},...,S_{m}$) from the stabilizer set $T$, the stabilizer set after the measurement can be obtained by replacing $S_{1}$ with $S^{*}$ and $S_{2},....S_{m}$ with $S_{2}S_{1},....S_{m}S_{1}$. From the prospective of the stabilizer matrix, this involves replacing the row corresponding to $S_{1}$ with the binary presentation of $S^{*}$ and adding the binary representation of $S_{1}$ to the rows corresponding to $S_{2},...,S_{m}$. 

Given a stabilizer matrix $M_{T}$, the entanglement entropy of any subsystem $A$ can be evaluated as
\begin{equation}
S_{A} = rank(M_{T}\big|_{A}) - L_{A}
\end{equation}
where $M_{T}\big|_{A}$ is the stabilizer matrix obtained by retaining only the columns corresponding to subsystem $A$, $L_{A}$ is the number of qubits in subsystem $A$. 

\subsection{Finite size scaling}
Next, we briefly mention the approach used to determine the values of critical points and critical exponents. We begin by assuming that our quantities of interest - the topological entanglement entropy $S_{topo}$ and mutual information $I(A,B)$ ($I_{AB}$) exhibit the following finite-size scaling forms:
\begin{equation}
S_{topo}(P,L) = F((P-P_{c})L^{\frac{1}{\nu}})
\end{equation}
and
\begin{equation}
|I_{AB}(P,L)-I_{AB}(P_{c},L)| = G((P-P_{c})L^{\frac{1}{\nu}})
\end{equation}
where $P_{c}$ is the critical probability and $\nu$ is the correlation length critical exponent. If we plot $S_{topo}$ as a function of $(P-P_{c})L^{\frac{1}{\nu}}$, we expect that all data points will collapse onto a single curve $F(x)$ for appropriate choices of $P_{c}$ and $\nu$. To find the optimal values of $P_{c}$ and $\nu$, we define the error function $\epsilon(P_{c},\nu)$ as follows:
\begin{equation}
\epsilon(P_{c},\nu) = \frac{1}{n-2}\sum_{k=2}^{n-1}(\tilde{y_{k}}-y_{k})^2
\label{Eq:error}
\end{equation}
where
\begin{equation}
\tilde{y_{k}} = \frac{(x_{k+1}-x_{k})y_{k-1}-(x_{k-1}-x_{k})y_{k+1}}{x_{k+1}-x_{k-1}}
\end{equation}
Here, $x_{k} = (P_{k}-P_{c})L_{k}^{\frac{1}{\nu}}$ and $y_{k}=S_{topo}(P_{k},L_{k})$. The index $k$ labels the $k^{th}$ data point, which includes system sizes $L=32,64,128$ and $256$, arranged in acsending order ($x_{1}<x_{2}<...<x_{n}$). The value $\tilde{y_{k}}$ represents the ideal $y_{k}$, calculated from the line passing through its neighboring points $(x_{k-1},y_{k-1})$ and $(x_{k+1},y_{k+1})$. Thus, the error function $\epsilon$ approaches zero if all data points collapse onto a single line.

\section{Numerical Results}
The numerical results for various entanglement measures presented in this section are obtained by averaging over $N=10^{4}$ circuit configurations, with the initial state set to $|+\rangle^{\otimes L}$. First and foremost, we need to examine whether different initial states influence our determination of critical points and phase boundaries.

\begin{figure}[htbp]
\begin{center}
\includegraphics[scale=0.41]{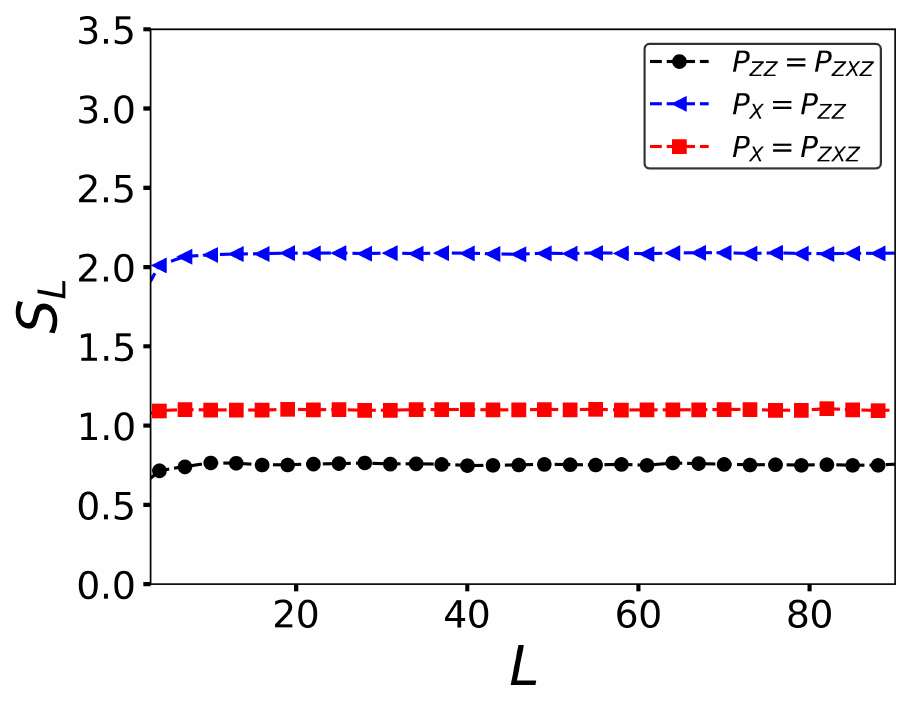}
\caption{The steady state Von Neumann entanglement entropy $S_{L}$ as a function of the size of subsystem $L$. The blue line correspond to $P_{ZXZ}=0.7$ with $P_{x}=P_{ZZ}$. The red line correspond to $P_{ZZ}=0.7$ with $P_{x}=P_{ZXZ}$. The black line correspond to $P_{X}=0.7$ with $P_{ZZ}=P_{ZXZ}$.}
\label{fig:area law}
\end{center}
\end{figure}

\subsection{Different initial states}
In Fig.~\ref{fig:EE initial state}, we observe how the half-chain Von Neumann entanglement entropy varies with time for two initial states $|+\rangle^{\otimes L}$ and $|0\rangle^{\otimes L}$. The parameters $P_{X}$,$P_{ZZ}$, and $P_{ZXZ}$ were positioned at the center of the phase diagram. The Von Neumann entanglement entropy for $|+\rangle^{\otimes L}$ approaches the equilibrium value faster than that for $|0\rangle^{\otimes L}$. However, both lines ultimately stabilize at the same plateau value. Therefore, we can conclude that our subsequent results are largely unaffected by different initial states. 

\begin{figure}[htbp]
\includegraphics[scale=0.46]{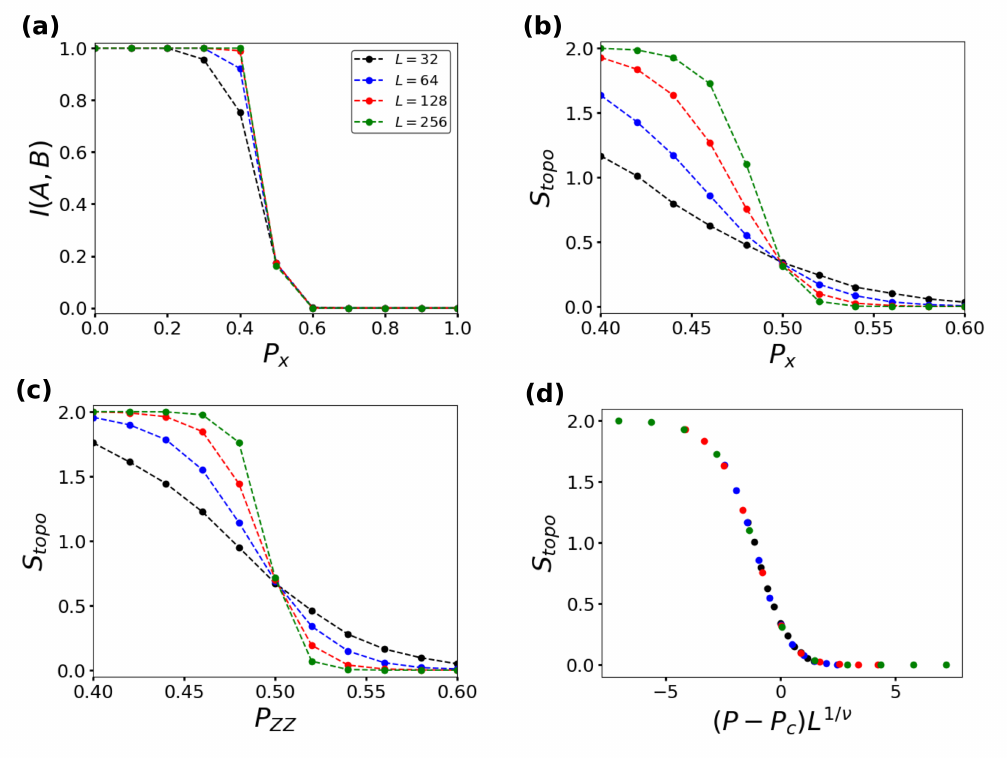}
\caption{Determination of three critical points and their critical exponents. \textbf{(a)}: The mutual information $I(A,B)$ versus the probability of single-qubit measurement $P_{X}$ with $P_{ZXZ}=0$. \textbf{(b)}: The topological entanglement entropy $S_{topo}$ versus the probability of single-qubit measurement $P_{X}$ with $P_{ZZ}=0$. \textbf{(c)}: The topological entanglement entropy $S_{topo}$ versus the probability of two-qubit measurement $P_{ZZ}$ with $P_{X}=0$. \textbf{(d)}: Scaling collapse of data from \textbf{(c)}, with $P_{c}=0.5$ and $\nu=1.3$. Scaling analysis for \textbf{(a)} and \textbf{(b)} are shown in the supplementary material.}
\label{fig:Critical points}
\end{figure}

\subsection{Area law}
Next, we investigate whether the system exhibits a volume law or area law. As illustrated in Fig.~\ref{fig:area law}, the steady state Von Neumann entanglement entropy remains insensitive to the size of subsystem $L$ as it varies. The subsystem is located at $[L_{tot}/2-L/2,L_{tot}/2+L/2]$, with $L_{tot}=128$. These parameters are chosen to be deep within three distinct phases. This result suggests that each phase supports an area law. A potential volume law can arise if two-qubit random unitary gates are incorporated into the circuit~\cite{ref38}.

\begin{figure}[htbp]
\includegraphics[scale=0.62]{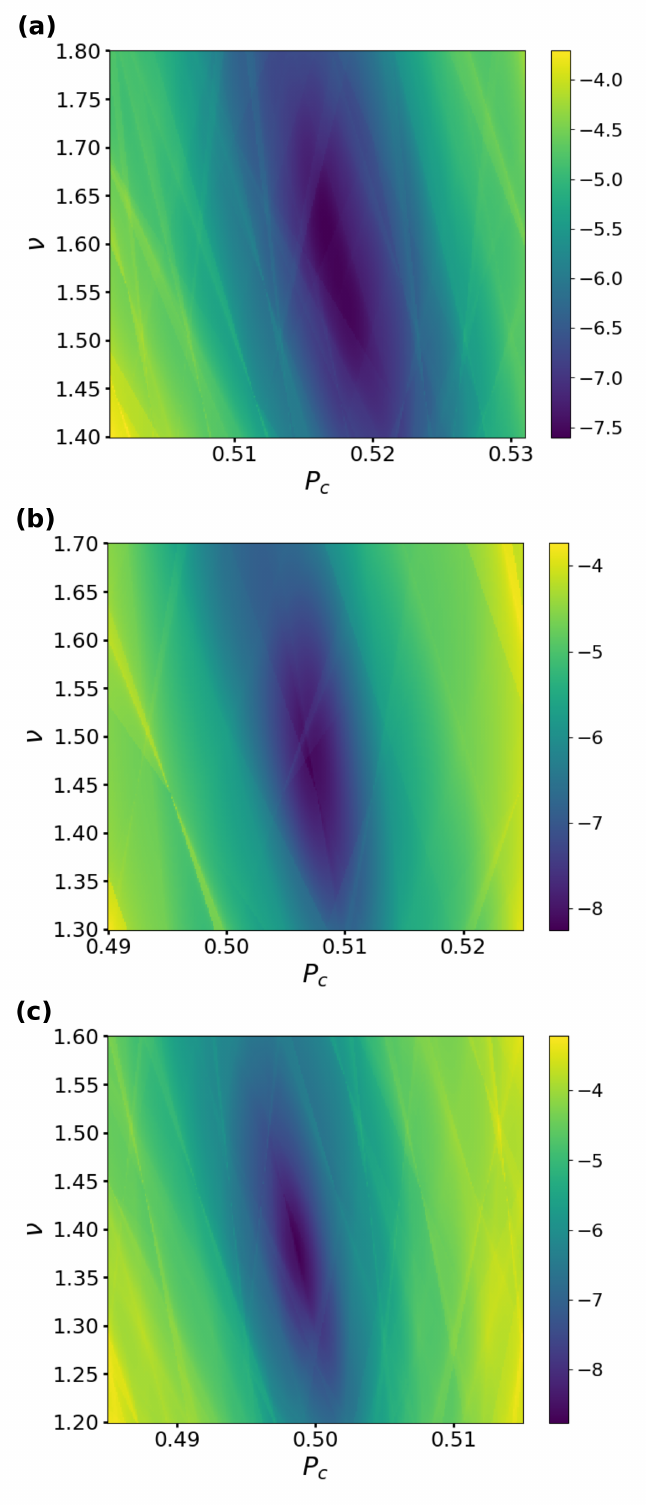}
\caption{Density plot of the error function $\epsilon(p_{c},\nu)$, as defined in Eq.~\ref{Eq:error}, with the color scale indicating the deviation from a perfect collapse. The parameters are chosen to correspond to the stars at the upper critical line in Fig.~\ref{fig:Phase Diagram}. Here $P_{c}$ stands for $P_{ZXZ}$. \textbf{(a):} $P_{X}=3P_{ZZ}$ ($P_{c}=0.517$, $\nu=1.58$). \textbf{(b):} $P_{X}=P_{ZZ}$ ($P_{c}=0.508$, $\nu=1.47$). \textbf{(c):} $P_{ZZ}=3P_{X}$ ($P_{c}=0.499$, $\nu=1.38$).}
\label{fig:upper critical}
\end{figure}

\subsection{Critical points}
We now focus on identifying the critical points that separate our system into distinct regions. This identification is characterized by changes in specific entanglement measures. As shown in Fig.~\ref{fig:Critical points}, we observe three critical points located at each boundary of the phase diagram.

In Fig.~\ref{fig:Critical points}(a), we plot the mutual information between two distant qubits as a function of the probability of a single-qubit measurement $P_{X}$ with $P_{ZXZ}$ set to zero. Both subsystems $A$ and $B$ are chosen to be single qubits, located at $L/8$ and $7L/8$. As $P_{X}$ varies, we observe that the mutual information decreases from one to zero, with all lines crossing around point $P_{X}=0.5$. This behavior can be easily understood by considering extreme cases. When $P_{X}=0$, the $ZZ$ measurement collapses the product state into an entangled state. For instance, $|++\rangle$ collapses into $|00\rangle+|11\rangle$ (ignoring the pre-factor) if the measurement outcome is $+1$. As the system reaches a steady state, the wavefunction eventually evolves into a GHZ (Greenberger-Horne-Zeilinger) state~\cite{ref46}, given by
\begin{equation}
|00...0\rangle+|11...1\rangle
\end{equation}
This state is indeed long-range entangled. By contrast, when $P_{X}=1$, the qubits remain in a product state with zero entanglement. These two behaviours are fundamentally different and cannot be analytically continued from one to the other, indicating the necessity of a phase transition between them.

\begin{figure}[htbp]
\begin{center}
\includegraphics[scale=0.62]{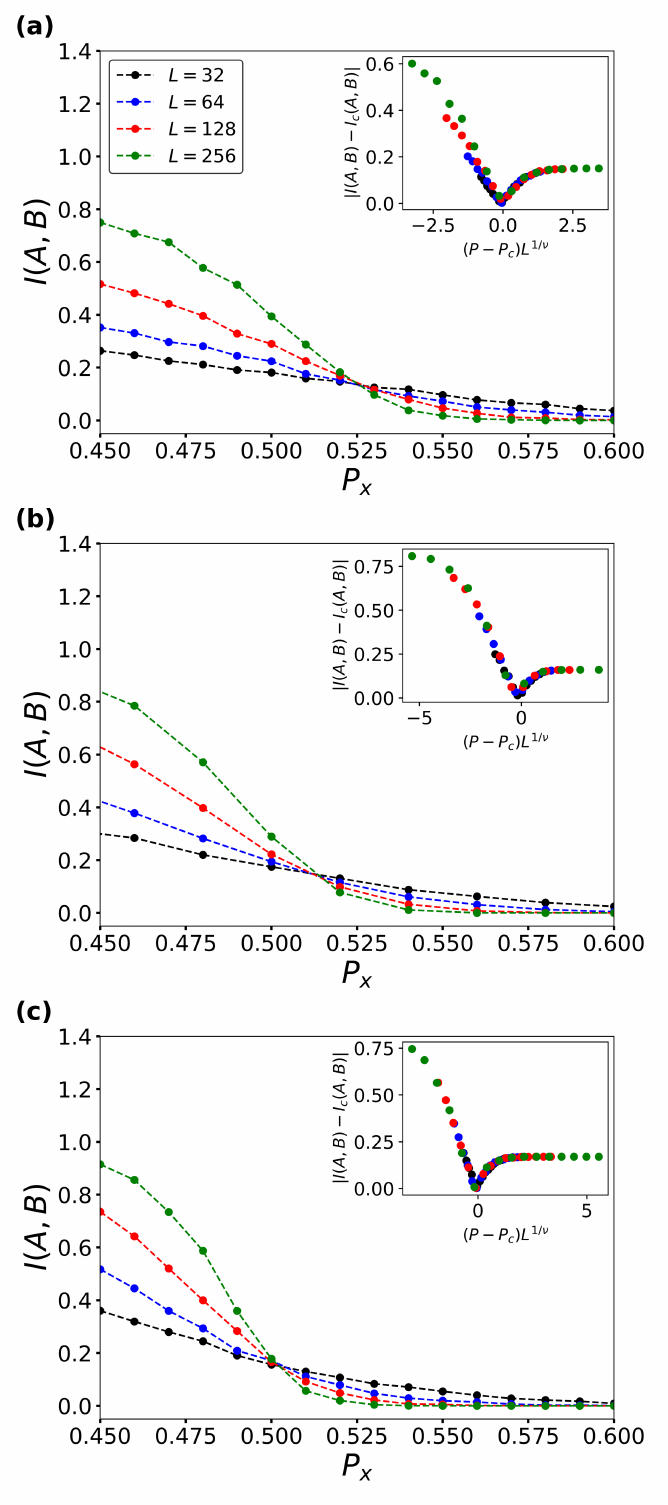}
\caption{The mutual information $I(A,B)$ between two distant spins as a function of the single-qubit measurement probability $P_{X}$ is illustrated, with different system sizes indicated by color: Black for $L=32$, Blue for $L=64$, Red for $L=128$, and Green for $L=256$. The inset shows the scaling analysis of the data from the main panel. The parameters are chosen to correspond to the stars at the lower critical line of the phase diagram. Here $P_{c}$ stands for $P_{X}$. \textbf{(a):} $P_{ZXZ}=3P_{ZZ}$ ($P_{c}=0.523$, $\nu=1.46$). \textbf{(b):} $P_{ZXZ}=P_{ZZ}$ ($P_{c}=0.517$, $\nu=1.45$). \textbf{(c):} $P_{ZZ}=3P_{ZXZ}$ ($P_{c}=0.502$, $\nu=1.37$). }
\label{fig:CL Lower}
\end{center}
\end{figure}

As shown in Fig.~\ref{fig:Critical points}(b), the topological entanglement entropy $S_{topo}$ is plotted as a function of the probability of a single-qubit measurement $P_{X}$ with $P_{ZZ}$ set to zero. We also observe that $S_{topo}$ decreases from a non-integer value to zero as $P_{X}$ increases, indicating a phase transition from a SPT phase to a topological trivial phase. This SPT phase ~\cite{ref47,ref48,ref49} is protected by a $Z_{2} \times Z_{2}$ symmetry, generated by $\prod_{even}X_{i}$ and $\prod_{odd}X_{i}$, and can also be realized as the ground state of the Hamilitonian:
\begin{equation}
H = -\sum_{i=2}^{L-1}Z_{i-1}X_{i}Z_{i+1}
\end{equation}
When $P_{X}=0$, the $ZXZ$ measurements dominate the circuit, resulting in an SPT phase with $S_{topo}=2$ in the thermodynamic limit ($L \rightarrow{\infty}$). Alternatively, a product state exhibits no topological order (i.e. $S_{topo}=0$) when only single qubit measurements are present in the circuit. Consequently, a critical point must lie between the two distinct phases. In addition, we also observe that all lines converge around $P_{X}=0.5$.

\begin{figure}[htbp]
\includegraphics[scale=0.6]{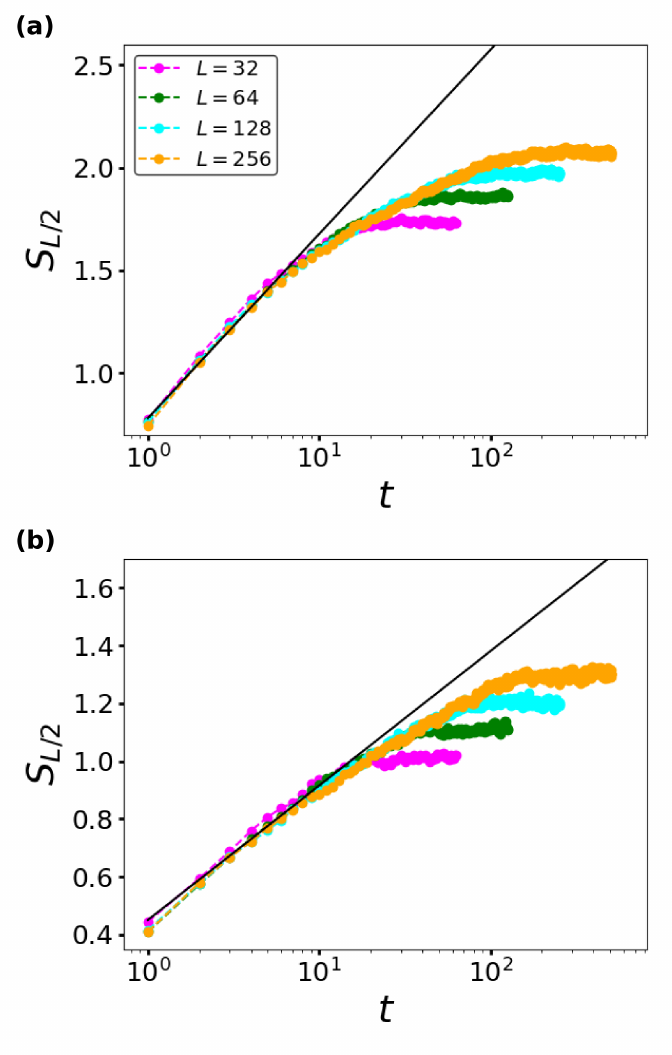}
\caption{The half-chain entanglement entropy $S_{topo}$ is plotted against $t=N/L^{z}$, where $N$ represents the total number of updating steps and the dynamical exponent $z$ set to $1$.  The parameters in each figure correspond to a critical point located on one of the critical lines. The black lines follow a fit to the form $S_{L/2}=a_{t}log(t)+b$. \textbf{(a):} $P_{X}=P_{ZZ}$, $P_{ZXZ}=0.508$, $a_{t}=0.27$, and $b=0.78$. \textbf{(b):} $P_{ZZ}=P_{ZXZ}$, $P_{X}=0.517$, $a_{t}=0.14$, and $b=0.45$.}
\label{fig:dynamical exponent}
\end{figure}

Fig.~\ref{fig:Critical points}(c) displays $S_{topo}$ as a function of the probability of two-qubit measurement $P_{ZZ}$ with $P_{X}=0$. As expected, a phase transition occurs as the measurement probability is varied. Additionally, we perform a scaling analysis, as introduced earlier, to determine the critical point $P_{c}$ and the correlation length critical exponent $\nu$ for Fig.~\ref{fig:Critical points}(a), (b), and (c). The data from different system sizes collapse onto a single curve when $P_{c}=0.5$ and $\nu=1.3$. The values of  the critical points can be understood through duality mapping between different measurement operators. Two duality transformations~\cite{ref48} exist between $X$ and $ZZ$, as well as between $X$ and $ZXZ$, defined as follows:
\begin{align}
U_{1}X_{i}U_{1}^{\dagger}=Z_{i}Z_{i+1} \\
U_{2}X_{i}U_{2}^{\dagger}=Z_{i-1}X_{i}Z_{i+1} 
\end{align}
Here, $U_{1}$ and $U_{2}$ are non-local unitary transformations. Furthermore, one can verify that $U_{2}U_{1}^{\dagger}$ maps $Z_{i}Z_{i+1}$ to $Z_{i-1}X_{i}Z_{i+1}$ which demonstrates that any two measurement operators can be related through a unitary transformation. Consequently, the critical points must coincide with the self-dual points ($P_{X}=P_{ZZ}$, $P_{X}=P_{ZXZ}$, and $P_{ZZ}=P_{ZXZ}$) as long as only two types of measurements are present in the circuit. The value of the critical exponent $\nu$ is close to that of a two dimensional percolation model, which is not a coincidence. This connection is elaborated in detail in the next section. 
\subsection{Critical lines}
Finally, we are prepared to examine the existence of critical lines and investigate the variation in several critical exponents along these lines. 

\begin{figure}[htbp]
\begin{center}
\includegraphics[scale=0.42]{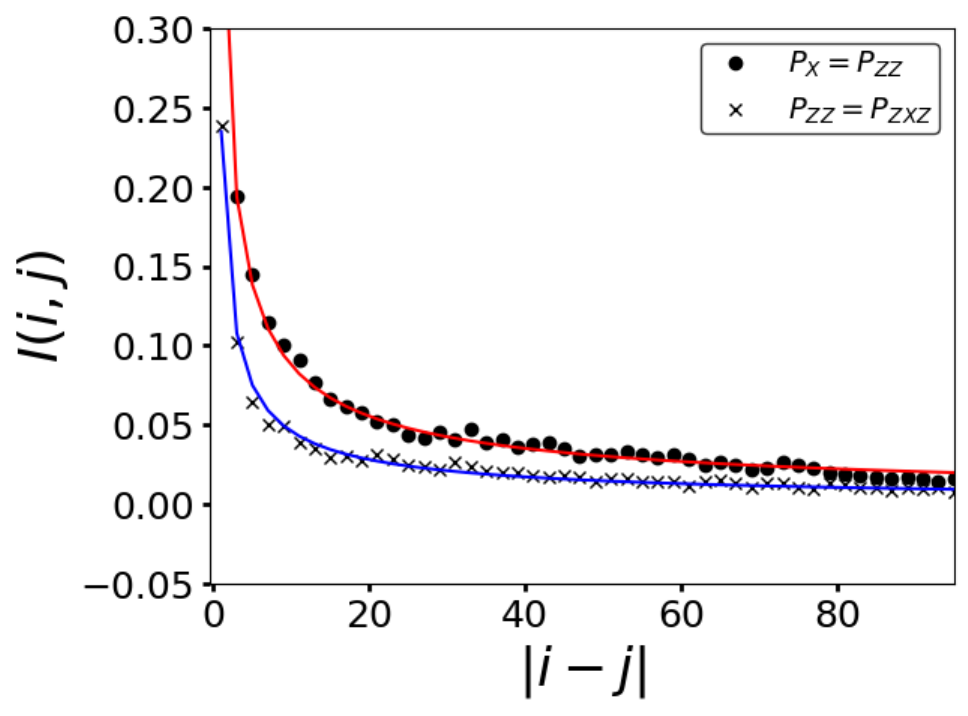}
\caption{The mutual information $I(i,j)$ between two qubits is plotted against their separation at two critical points with system size $L=128$. Black dots represent the case where $P_{X}=P_{ZZ}$ and $P_{ZXZ}=0$, while black crosses correspond to $P_{ZZ}=P_{ZXZ}$ and $P_{X}=0.517$. The blue and red lines are fits of the form $\propto1/|i-j|^{K}$, with blue for $K=0.71$ and red for $K=0.66$.}
\label{fig:separation pl}
\end{center}
\end{figure}

To pinpoint each critical line, we employ a specific entanglement measure and explore three distinct paths in the parameter space. Each path has a different constraint, but shares a common variable. For the upper (lower) critical line, we use $S_{topo}$ ($I(A,B)$) and vary $P_{ZXZ}$ ($P_{X})$, while keeping $P_{X}=3P_{ZZ}$ ($P_{ZXZ}=3P_{ZZ}$), $P_{X}=P_{ZZ}$ ($P_{ZXZ}=P_{ZZ}$), and $P_{X}=1/3P_{ZZ}$ ($P_{ZXZ}=1/3P_{ZZ}$) along three distinct paths. The error functions $\epsilon$ for upper critical line are illustrated in Fig.~\ref{fig:upper critical}. We find that the critical line is not a straight line connecting two critical points, as $P_{c}$ varies with $P_{X}$. Additionally, the critical exponent $\nu$ deviates from its original value of $1.3$, as observed at the boundary of the phase diagram. This suggests that the standard percolation model is inadequate to describe these critical points. A similar behavior is observed for the lower critical line, as shown in Fig.~\ref{fig:CL Lower}. Interestingly, there is no critical line connecting the two critical points ($P_{X}=P_{ZZ}$, $P_{ZXZ}=0$ and $P_{ZZ}=P_{ZXZ}$, $P_{X}=0$) as the mutual information changes continuously in that area. 

In Fig.~\ref{fig:dynamical exponent}, we demonstrate that the growth of theentanglement entropy at early times follows the desired logarithmic form $a_{t}log(t/L^{z})+b$ ~\cite{ref50} at two critical points, each selected from a different critical line. In addition, we observe that the entanglement entropy across different system sizes collapses onto this logarithmic function when the dynamical exponent $z$ is set to one, until the finite-size effects become noticeable. This behaviour indicates that the physics underlying these critical points exhibits Lorentz invariance and is described by conformal field theory (CFT). However, unlike in unitary CFTs, the coefficient of $a_{t}$ we find is not determined by the central charge of the CFT. Instead, its interpretation ~\cite{ref14} is only known in the context of the volume law to area law transition. 

Finally, we focus on how mutual information $I(i,j)$ changes with the distance $d=|i-j|$ between two qubits along the lower critical line. Qubits $i$ and $j$ are positioned at $L/2-d/2$ and $L/2+d/2$ respectively. As shown in Fig.~\ref{fig:separation pl}, $I(i,j)$ for distant qubits attains a finite value at the entanglement phase transition. This value follows a power-law decay with respect to distance $d$, characterized by a critical exponent $K$. This behavior aligns with our expectations, stemming from the scale invariance property at the critical point. Similar to the critical exponent $\nu$, $K$ changes as the measurement probability varies. 

\begin{figure}[htbp]
\begin{center}
\includegraphics[scale=0.59]{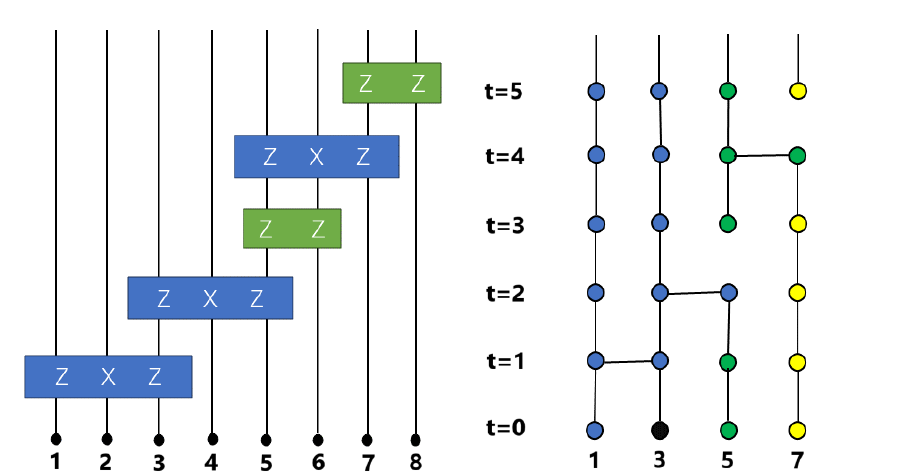}
\caption{The mapping from the quantum circuit to bond percolation is shown in detail, with time labeled in the middle. \textbf{Left}: A schematic drawing of a circuit configuration ($P_{X}=0$) with $8$ qubits, showing up to 5 updating steps. \textbf{Right}: Step-by-step evolution of the qubits from the odd sites in the bond percolation picture. Qubits within the same SPT cluster are connected by horizontal bars and shown in the same color. A similar diagram exists for the even sites. Together, these two pictures fully characterize the entanglement structure of the state.}
\label{fig:bond percolation}
\end{center}
\end{figure}

\section{Percolation Theory}
As previously mentioned, our $1+1$ dimensional circuit can be mapped to a two-dimensional bond percolation model when only two types of measurements are involved. in this section, we provide further details on this mapping. For illustrative purposes, we set the probability of single qubit measurements to zero and focus exclusively on the evolution of qubits located at the odd-numbered lattice sites. 

Initially, each qubit is assigned a distinct color, indicating that the state is a product state with no correlations between qubits. As the system evolves, if a $ZXZ$ measurement is applied to a group of qubits, these qubits are assigned the same color, indicating that they belong to the same SPT cluster. In this case, a horiztontal bond is drawn between the qubits. Additionally, a vertical bond is drawn between successive updating steps of each qubit, unless a $ZZ$ measurement is applied to that qubit. If a qubit is already part of a SPT cluster and a $ZZ$ measurement is applied. The qubit is removed from the cluster, assigned a different color and no vertical bond is drawn for that updating step. Those rules are illustrated in Fig.~\ref{fig:bond percolation}. For instance, a horizontal bond is drawn between qubits $3$ and $5$ since a $ZXZ$ measurement is applied to qubits $3$, $4$, and $5$. Meanwhile, the vertical bond for the qubit $5$ is absent between $t=2$ and $t=3$ due to a $ZZ$ measurement applied at $t=3$. 

In the two-dimensional bond percolation model, the critical point occurs when the probability of a bond being occupied is $0.5$ for a square lattice, and the critical exponent $\nu$ is $4/3$~\cite{ref51}. In our model, the probability of drawing a horizontal bond is $\frac{1}{2}P_{ZXZ}$, and the probability of drawing a vertical bond
is $\frac{1}{2}(1-P_{ZZ})$. Because both probabilities are equal, criticality occurs when $\frac{1}{2}P_{ZXZ}$ equals $0.25$ (as we only consider the odd-numbered sites). A similar argument can be made for cases where $P_{ZXZ}=0$ and $P_{ZZ}=0$. Thus, the conformal theory described by these critical points falls into the same universality class as the bond percolation model.

However, this type of mapping fails to be established in the interior of the phase diagram, as the values of the critical exponents we verified deviate from those of the percolation model.
 
\section{Summary and Discussion}
In this study, we introduce and explore a measurement-only model that involves three types of measurement gates. Owing to the non-commutativity of any two measurement operators, the model displays two critical lines that separate the three distinct phases, each characterized by a unique entanglement structure. Through scaling analysis, we find that critical exponents vary along these critical lines. We establish a relationship between our model and a two-dimensional bond percolation model by verifying the locations of the critical points and the values of the critical exponents within certain limits. Notably, we observe that the location of the critical point we obtained aligns with that derived from the Hamiltonian's picture in Eq.~\ref{Eq:Hamilitonian} when only two measurement operators are present. For example, if we relate the coupling parameters $h$ and $\lambda_{1}$ in Eq.~\ref{Eq:Hamilitonian} to the measurement probabilities $P_{X}$ and $P_{ZZ}$, the condition for criticality in the quantum Ising model, $h=\lambda_{1}$, is transformed into $P_{X}=P_{ZZ}$. A similar argument holds true for $P_{X}$ and $P_{ZXZ}$. Thus, in this case, the state obtained through repeated projective measurements exhibits properties analogous to the ground state of the Hamiltonian. However, this relationship does not generally hold when all the measurements are present in the circuit. 

\section*{ACKNOWLEDGMENTS}
The authors would like to thank Shi-Xin Zhang for helpful discussions and suggestions on the manuscript. The work is  supported by the Ministry of Science and Technology  (Grant No. 2022YFA1403900), the Strategic Priority Research Program of the Chinese Academy of Sciences (Grant No. XDB28000000, XDB33000000), the New Cornerstone Investigator Program, and the International Young Scientist Fellowship of Institute of Physics Chinese Academy of Science (No.202308).

\section*{Data availability.}  Numerical data for this manuscript are publicly accessible in Ref. \cite{data-availbale}.



\bibliography{ref.bib}

\end{document}